# Climate Change Adaptation in the British Columbia Wine Industry

## Can carbon sequestration technology lower the B.C. Wine Industry's greenhouse gas emissions?

Lee Cartier[1] and Svan Lembke

Okanagan School of Business, Okanagan College

## Abstract

*Purpose*

The purpose of this study is to measure the benefits and costs of using biochar, a carbon sequestration technology output, to reduce the B.C Wine Industry's carbon emissions. In this study, carbon sequestration refers to the process of capturing and storing atmospheric carbon dioxide in the soil. Opportunities for business innovation and development, and sustainable wine production processes in the Okanagan valley are identified and discussed.

*Design/methodology/approach*

An economic model was developed to calculate the value-added for each of the three sectors that comprise the BC Wine industry: first, grape growing with biochar (vineyard sector), second, wine production and sales with grapes grown with biochar (winery sector), and third, biomass collection and biochar production (biochar sector). Two value chain scenarios were developed: one with an independent biochar production sector, and a second with an integrated biochar production sector. At the time of this study no primary data related to the economics of biochar use in the BC wine industry exists. The assumptions for each variable used in this study are drawn from the literature and prior research by the authors. Four analytical approaches are used in the study: benefit-cost analysis, Monte Carlo analysis, sensitivity analysis, and $CO_2$ sequestration cost.

*Findings*

Biochar produced from wine industry waste was applied to 288 Hectares of vineyard, the maximum waste that can be collected in the BC wine industry. Each sector of the wine value chain is potentially profitable. The mean value chain NPV for the integrated biochar scenario is $19.3 million: $13.2 million for the biochar production segment, $864,725 for the vineyard segment, and $5.2 million for the winery segment. The probability of exceeding the break-even point (Benefit Cost ratio >1.0) is 99% for the biochar segment, 93% for the vineyard segment, and 100% for the winery segment. The amount of carbon that can be sequestered each year from wine industry waste is 9,000 tonnes, this is equivalent of removing 2,200 automobiles from the road.

The implications for producing biochar as a profitable independent business are likely minimal compared to what could be achieved along the value chain. Biochar use as a soil amendment is a long-term investment for farmers with results best assessed after multiple years. Today's practices for grape growing and wine making are based primarily on grape quantity per acre, not the quality-quantity ratio of the grapes produced. Some re-evaluations of these practices in light of long-term biochar effects could transform the profitability and market positioning of the industry.

---

[1] lcartier@okanagan.bc.ca

*Practical implications*

Results from this study indicate that biochar production from wine industry waste and use in the vineyard can make a significant economic contribution to the BC wine industry, and enhance its Sustainable Winegrowing BC program, specifically the Climate Action standard related to reducing GHG emissions. Recommendations include using a biochar strategy to differentiate itself from imports by branding its wines as 'climate friendly'. Biochar adoption in new vineyard plantings is also recommended. The scalability and options for adoption in the BC wine industry are important subsequent research topics to enable the industry to implement an integrated approach and stay globally competitive.

*Originality/value*

This study is unique as it investigates the value to the BC wine industry of using biochar, produced from wine industry waste, to sequester atmospheric $CO_2$. It demonstrates that using biochar to sequester atmospheric $CO_2$ can be profitable.

# Introduction

The overwhelming body of scientific evidence shows that global temperatures due to CO2 emissions from human industrial activity are increasing, and the rising temperature will have a significant impact on Canadian agriculture (USGCRP, 2017; Bush and Lemmen, 2019; Warren & Lemmen, 2014). Moreover, Canadian agricultural activity makes a significant contribution to Canada's greenhouse gas emissions (Pembina Institute, 2010).

Governments are beginning to respond to the impact of climate change. The Canadian government is committed to becoming carbon neutral by 2050. This will involve developing programs that reduce $CO_2$ emissions, using carbon offsets to compensate for ongoing $CO_2$ emission activities, and investing in sequestration technologies that capture CO2 before it is released into the atmosphere (Government of Canada, nda). In British Columbia (BC), the provincial government has set targets to reduce GHG by 40% by 2030 and 80% by 2050. To help achieve these targets the government had purchased 5.6 million tonnes of carbon offsets by 2017 (United Nations, nd).

Canadian business is also moving to embrace carbon neutrality, even in its most carbon intensive oil and gas industries. These companies are adopting new processes and technology to reduce their carbon footprint. For example, Suncor Energy will invest $1.4 billion to construct a new power cogeneration plan that will remove 2.5 megatons of $CO_2$ from their operations, and Shell Canada has built a carbon capture and storage facility that removes 4 million tonnes of $CO_2$. Overall, Alberta oil and gas companies have reduced their GHG emissions by 28% per barrel of oil produced since 2000 (Government of Canada, ndb).

We cannot reverse the impact of climate change, but we can take steps to adapt to these changes. The purpose of this study is to measure the benefits and costs of using carbon sequestration technology to reduce the B.C Wine Industry's carbon emissions. In this study, carbon sequestration refers to the process of capturing and storing atmospheric carbon dioxide in the soil[2].

## Climate Change Adaptation in the BC Wine Industry

In British Columbia, the BC wine industry is taking a leadership role in climate change adaptation. Through its Sustainable Winegrowing BC program (SWBC), the industry is taking an aggressive approach to mitigating climate change. The industry has established two sustainability standards: the SWBC Vineyard Standard and the SWBC Wineries Standard. These standards form the basis of sustainability certification. The wineries certification program identifies seven standards:

1. Setting the sustainability foundation
2. Water efficiency and conservation
3. Energy efficiency and management
4. Responsible waste management
5. Climate action
    a. Reduced GHG emissions
    b. Safe and reduced use of hazardous substances
    c. Preparation for disasters and extreme weather events
6. Social equity
7. Eco-efficient and sustainable winery infrastructure (Sustainable Wineries SWBC Standard)

This research project is focused on the Climate Action standard, specifically reducing GHG emissions.

## Methods to Mitigate Climate Change

According to Budinis (2020), "achieving carbon neutrality, or "net zero," means that any $CO_2$ released into the atmosphere from human activity is offset by an equivalent amount being removed." This means that achieving net-zero will require more than just reducing human $CO_2$ emissions. It will also require the use of negative emissions technology (NETs) to remove existing $CO_2$ from the atmosphere (Gasser, Guivarch, Tachiiri, Jones & Ciais, 2015).

Globally, signatories to the Paris Climate Agreement are rapidly moving to reduce their $CO_2$ emissions by replacing fossil fuel energy sources with renewable sources such as solar, wind power and hydrogen, and electrifying their transportation systems. Although the global adoption of NETs has been slower than for emissions reduction technology, these technologies already exist.

## Negative Emissions Technology

The focus of this study is the removal and sequestration of $CO_2$. Further, the analysis is based on the premise that useful products, with economic value, can be created from atmospheric $CO_2$. Hepburn, Adlen, Beddington, Carter, Fuss, Mac Dowell, Minx, Smith & Williams (2019) discuss ten different pathways to create useful products for atmospheric $CO_2$:

---

[2] Source: U.S. Geological Survey. What is carbon sequestration?
Retrieved from https://www.usgs.gov/faqs/what-carbon-sequestration?qt-news_science_products=0#qt-news_science_products

1. Chemicals from $CO_2$
2. Fuels from $CO_2$
3. Products from microalgae
4. Concrete building materials
5. $CO_2$ - EOR (Enhanced Oil Recovery)
6. Bioenergy with carbon capture and storage
7. Enhanced weathering
8. Forestry techniques
9. Soil carbon sequestration techniques
10. Biochar

A summary of each pathway is reproduced in Appendix A. Biochar production has been chosen for this study. Biochar will be produced from wine industry biomass waste (pomace and grape prunings) and then used as a soil amendment in the vineyard. The result is that atmospheric $CO_2$ removed through photosynthesis is sequestered in the vineyard soils. Biochar is a form of charcoal and is very stable. It can remain in the soil for hundreds or thousands of years.

## How is Biochar Produced?

Biochar is produced through the process of pyrolysis. Pyrolysis involves heating biomass in the absence of air or oxygen. Once the biomass is fed into the pyrolizer and ignited oxygen is removed from the chamber. Once initiated the process is exothermic and continues without further energy input until all the biomass is converted into finished products. Pyrolysis has the potential to produce three products: biochar, bio-oil, and pyrolysis gas (syngas), all of which have commercial value. The mix of products is controlled by adjusting the temperature in the pyrolizer. Under high temperatures (fast pyrolysis) all three products are produced. Under lower temperatures (slow pyrolysis) most of the biomass is converted to biochar and pyrolysis gas with the gas used as an internal energy source to maintain the process.

This feasibility study is based on the use of slow pyrolysis to produce biochar and is represented in Figure 1.

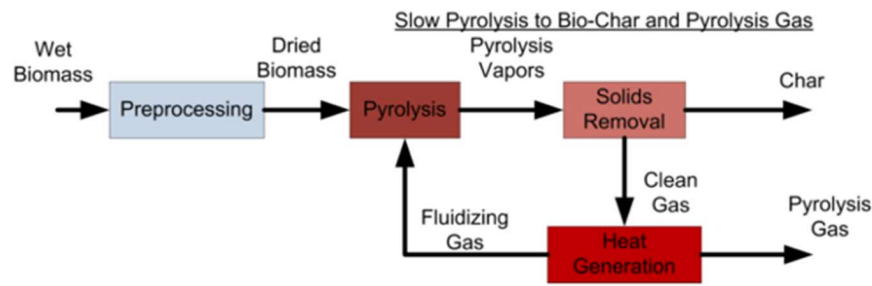

Figure 1: Slow Pyrolysis Process[3].

---
[3] Source: Brown, T., Wright, M, & Brown, R. (nd)

## What We Know About Biochar

There are many benefits associated with biochar use on vineyard soils. When used as a soil amendment, biochar can help remediate contaminated agricultural soil. It can help improve soil fertility by reducing its acidity - improving nutrient availability (C, N, Ca, Mg, K, and P). It can help soil overcome biotic stress, remove heavy-metal pollutants, and increase the ability to retain moisture, helping to attract more useful fungi and other microbes while controlling pathogens. Biochar can also suppress greenhouse gases in soil, specifically emissions of methane and nitrous oxide (Rawat, Saxena & Sanwal, 2019).

Biochar use can increase crop yields. Timmons, Lema-Driscoll & Gazi Uddin (2017) reported that in one study tomatoes experienced a 20% crop yield increase when biochar was mixed with fertilizer, compared to fertilizer only. In another study with peppers, it was found that biochar application could raise whole plant yield up to 66.4% against the control plant. Keske (2020) reports that biochar applications increase potato yields approximately 19% based upon a meta-analysis of 59 pot experiments from 21 countries and 57 field experiments from 21 countries. In another study they, found a 28.6% average increase in vegetable crop yields.

Biochar use in young apple plantings can promote plant growth at early growth stages of apple orchards. Khorram, Zhang, Fatemi, Kiefer, Maddah, Baqar, Zakaria & Li (2019) found that the trunk diameter and shoot number of apple trees increased 23–26% by the end of the first year. However, they did not find any increase in apple yield.

When used in vineyards, biochar application can improve the resilience of vineyards against drought and improve grape yields (Baronti, Vaccari, Miglietta, Calzolari, Lugato, Orlandini, Pini, Zulian & Genesio, 2014). Biochar can improve grapevine fine root development. Amendola, Montagnoli, Terzaghi, Trupiano, Oliva, Baronti, Miglietta, Chiatante, & Scippa, (2017 showed that after biochar application there was a significant increase in fine root biomass, annual production, and vine lifespan.

Biochar use can increase grape yields. A four-year longitudinal study by Genesio, Miglietta, Baronti, & Vaccari (2015) showed higher yields, up to 66%, of treated plots with respect to their controls, while no significant differences were observed in grape quality parameters. While this appears promising, more research with grape yields are needed. Keske (2020) reported positive results for both beets and potato production. However other studies reported mixed results. Dickenson et.al. reported that biochar application for cereals agriculture in North West Europe was never positive, however; biochar application for cereals production in sub-Saharan Africa was profitable. This implies that biochar effectiveness may be strongly affected not only by the crop type but also by other variables such as the composition of the soil and other external factors.

In Canada, biochar use is regulated under the Fertilizer Act as a level II soil amendment. The Canadian Food Inspection Agency (CFIA) requires that all biochar sold as a soil amendment be certified. Certification requires that the biochar be tested at a certified testing laboratory (Government of Canada, ndc)

## What We Do Not Know About Biochar

Although a significant amount of research has been done on use of biochar in agriculture, there are major gaps in the economic information about biochar production and use. The inputs to pyrolysis are known but there is little agreement on the quantity and cost of these inputs.

Depending on how biochar is used the prices vary dramatically, from $120 to $3,000/Tonne for agriculture use. However, when sold through garden centres for home gardener use prices of $16,000/Tonne are identified.

There is no agreement on biochar application rates. Rates reported in the literature vary from 5 – 20 Tonnes/Hectare. Biochar can be combined with compost, but this does not seem to decrease the amount of biochar required per hectare.

More research is needed into the effect of biochar on grape yields. Although it is reported that grape yields can be increased, the reported increases range from 20%- 66%.

How frequently, if ever, biochar needs to be reapplied to the soil is unknown. Although it is very long lived in the soil (hundreds of years), it is unclear whether the effectiveness of biochar in the soil diminishes over time. There is some consensus that biochar does not have to be applied each year, but no estimates of frequency are identified in the literature.

## Conceptual Framework for this Study

The conceptual framework is based on two aspects: the boundaries of the study, and the approach to the analysis.

### Boundaries of the Study

The boundaries for this study are established by the circular economy model.

A linear (conventional) economy is based on the use of virgin finite resources in production: make-use-dispose. It is ultimately inefficient as much of the products value is lost when the item is disposed. Additional inefficiencies occur from the negative externalities associated from both the resource extraction process and ultimate disposal through landfills or incineration. The McKinsey Centre for Business and Environment (nd) calculates that in Europe, 60% of materials are either landfilled or incinerated and 40% is recycled. They conclude that 95% of the material and energy value is lost through disposal while only 5% is salvaged through recycling.

*Circular Economy Model*

In contrast to a linear economy, the circular economy is modeled on the metabolic processes found in natural ecosystems where the waste produced by one organism become an input for another one. These metabolic processes can act as templates for human industrial activity (Ellen MacArthur Foundation, nd). So rather than 'make-use-dispose', an industry can 'reuse, repair, remanufacture or recycle'. It utilizes the feedback loops that are already present in BC and applies them to human activity. The goal is to ultimately eliminate waste (within the limitations of the laws of thermodynamics). The model distinguishes between technical cycles (production of manufactured goods) and biological cycles (production of food). It seeks to apply the natural regenerative processes associated with biological cycles to manufacturing processes associated with the production of goods such as electronics and automobiles (Ellen MacArthur Foundation, nd).

In biological cycles, McKinsey (nd) estimates that the use of precision agriculture could reduce machinery and input costs by as much as 75%, a very appealing prospect for any region with an economy relying on agriculture.

*The Circular Economy in the Okanagan Region of British Columbia*

In the Okanagan, the agricultural products cluster is composed of two main value chains: the wine industry value chain and the tree fruit industry value chain. These two value chains draw on a common pool of resources such as land, water, labour, and capital (the factors of production). In the wine industry the value chain links involve grape growing activities, wine making activities, and marketing and sale activities.

There is waste associated with each link in the wine industry's value chain. In the wine industry vineyards grow a new grape vine canopy each year. At the end of the growing season, this canopy is removed through pruning and the vegetative matter is disposed of, either by mulching or burning. Wine making involves grape crushing, fermentation, and bottling activities. The waste from the crushed grapes (pomace) is disposed of in landfills or used in composting.

The waste disposal associated with these production activities, and the energy consumed during the manufacturing and sales activities produce $CO_2$ and other greenhouse gases which contributes to global warming and climate change.

*Creating a New Industry Value Chain*

There is an opportunity to create a new value chain using the waste associated with the production activities in vineyards and wineries. This new 'wine and biochar value chain' involves using the vegetative waste associated with production, prunings and pomace, as a feedstock to produce biochar. The biochar is then used as a soil amendment and conditioner that sequester $CO_2$ in the soil. This new value chain is represented in Figure 2.

*Value Chain Activities*

The first link in the value chain involves collecting the chipped canopy waste and pomace from vineyards and wineries. The biochar producer will pay producers for this material which creates value. The biochar is then sold to the grape growers and used as a soil amendment to improve soil quality and grape yields. The additional grapes are then sold to the wineries to increase their production and sale of wine.

In this study, the value chain is considered to be a closed system. That is, biochar production is limited to the quantity of prunings and pomace produced.; no outside feedstocks are imported. To capture the value added associated with the biochar production and use, all transactions are at market prices. Three benefits are identified: atmospheric $CO_2$ is captured and stored, grape yields are increased, and wine production is increased.

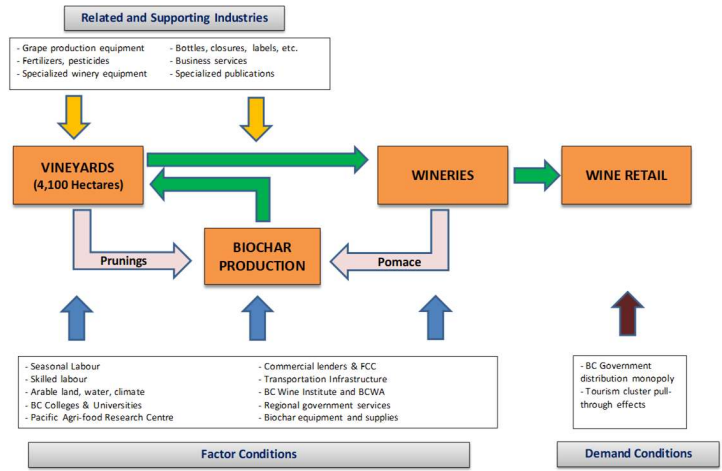

Figure 2: Wine and Biochar Value Chain

Approach to the Analysis

A proof-of concept approach is used to measure the value added associated with introducing a biochar link in the wine industry value chain. To pass the proof-of-concept, the concept must pass three tests: 1. the market test, 2. the technical feasibility test, and 3. the economic feasibility test. The market test passes as the size of the biochar market is bounded by the BC grape acreage (4,100 Hectares). The technical feasibility test passes as the pyrolysis technology required to make biochar is readily available. The economic feasibility is assessed using a benefit-cost analysis.

A Benefit - Cost analysis (B/C) is a standard method used to determine the economic viability of a project. Dickinson, Balduccio, Buysse, Ronsse, Van Huylenbroeck & Prins (2015) used this method to evaluate the economic viability of using biochar to improve cereal crops. As the name implies, it is prepared by dividing the present value of cash benefits derived from the project by the present value of the cash inputs required by the project. A project is considered potentially viable if the Benefit/Cost Ratio is greater than 1.0. Any project with a B/C ratio of less than 1.0 is considered uneconomic and would be abandoned. The analysis also provides an effective way to identify the most critical variables and measure the sensitivity of the B/C to these variables.

For this study, two value chain scenarios are developed:

1. *An independent biochar production sector (Independent Sector)*. In this scenario an independent producer purchases biomass from wineries and grape growers and sells the biochar produced to the grape growers. The biochar producer is a profit seeking enterprise.
2. *An integrated biochar production sector (Integrated Sector)*. In this scenario biochar production is integrated into the winery/vineyard operations. It operates as division of the winery and supplies biochar for both its own use in the vineyard as well as selling biochar to other grape growers.

### Research Question and Objectives

The research question guiding this investigation is "What are the benefits and costs of implementing a carbon sequestration strategy in the B.C. wine industry?" The four research objectives (RO) developed to answer this question are listed below.

> RO1: What are the benefits and costs of producing biochar from wine industry waste?
> RO2: What are the benefits and costs of using biochar as a soil amendment in the vineyard?
> RO3: What is the impact on the winery of using biochar in the vineyard?
> RO4: How much atmospheric $CO_2$ can be sequestered by producing biochar and using it as a soil amendment in the vineyard?

## Methodology

This section describes the four analytical approaches used in the study: the benefit-cost analysis, Monte Carlo analysis, sensitivity analysis, and $CO_2$ sequestration cost.

### Benefit-Cost Analysis

A B/C analysis is prepared for each sector in the value chain: biochar manufacturers, vineyards (grape growers), and wineries. The projected cash flows for each sector are discounted over a ten-year horizon. A discount rate of 10% is used for each sector. This discount rate is consistent with those used in other biochar economic analysis. Discount rates range from 7% (Haeldermansa, Campionc, Kuppensc, Vanreppelena, Cuypersd, & Schreurs, 2020) to 16.5% (Sahoo, Bilek, Bergman, & Mani, 2019). Campbell, Anderson, Daugaard, & Naughton (2018) used a 10% discount rate as their base rate, with an uncertainty distribution rate of 4% minimum rate and a 16% maximum rate.

At the time of this study no primary data related to the economics of biochar use in the BC wine industry exists. The assumptions used for each aspect of this study are drawn from the literature and prior research by the authors. Because these assumptions combine data from several different sources there is a high level of uncertainty associated with this data. The greater the uncertainty associated with these assumptions, the greater is the risk, or probability, that the outcome, the base B/C ratio, will not be achieved. The statistical approaches used to deal with this uncertainty is Monte Carlo Analysis and sensitivity analysis.

### Data sources

The data sources for the variables used in each sector are provided in Table 1. All prices and cost data used in this study are converted to Canadian dollars and adjusted for inflation.

Table 1: Data Sources

**Biochar Manufacturers Sector**

| Variable | Sources |
|---|---|
| Biochar prices | Campbell et.al. (2018) |
| | Bushell, (2018) |
| | Rogue Biochar - Oregon Biochar Solutions (A biochar retailer) |
| Biochar production costs | Campbell et.al. (2018) |
| | Keske, (2020) |
| | Haeldermansa et.al. (2020) |
| | Sahoo et.al. (2019) |
| Biochar application rates | Keske, (2020) |
| | Amendola et.al. (2017) |
| | Baronti et.al. (2014) |
| | Genesio et.al. (2015) |
| | Giagnoni, et.al. (2019). |
| | Khorram et. al. (2019) |
| | Timmons et.al. (2017) |
| Application frequency | Keske, (2020) |
| | Major (2010) |
| | Farm Folk City Folk (nd) |
| Grape pomace & prunings supply | Hogervorst, Miljić & Puškaš (2017) |
| | Skinkis, (2013) |
| Biochar conversion rates | Campbell et.al. (2018) |
| | Brown et.al. (nd) |

**Vineyard and Winery Sectors**

| Variable | Sources |
|---|---|
| Grape Prices | BC Wine Grape Council - Annual Crop Assessment - 2019 |
| Grape yields | Cartier, (2017) |
| Grape production cost | BC Ministry of Agriculture (2014) |
| | Cartier, (2017) |
| Grape acreage | B.C. Wine Grape Acreage Report (2014) Available from the BC Wine institute |
| | Cartier (2017) |
| Wine prices | BC Liquor Distribution Branch - Quarterly Market Reviews |
| | Cartier (2017) |
| Wine costs | Cartier (2017) |

## Monte Carlo Analysis

Monte Carlo is a mathematical simulation that calculates the probability of different outcomes based on multiple iterations of the model. The Monte Carlo probability analysis is well understood and is used to complete an economic analysis in several biochar studies (Campbell et. al., (2018); Dickinson et. al. (2015); Haeldermansa et.al. (2020)). The analysis establishes a range of estimates: minimum (most pessimistic), base (most likely), and maximum (most optimistic) for each variable. It then performs multiple iterations of the model by taking a random sample from each distribution during each iteration. When enough iterations are completed, the probability that the expected outcome will be achieved is established. The range of values used for each variable was developed from the data sources listed above. The uncertainty distributions associated with each variable are provided in Appendix B. In this study, 1,000 iterations were used to achieve a stable B/C probability distribution. An example of the uncertainty distribution for biochar price and biochar application rate is provided in Figure 3. Separate Monte Carlo simulations were completed for each of the two scenarios.

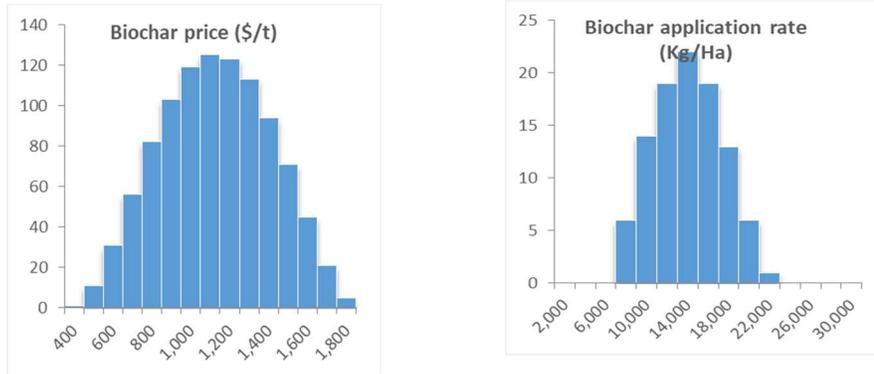

Figure 3: Biochar Price Distribution

Sensitivity Analysis

A sensitivity analysis was completed for each sector in the value chain. Regression analysis is used to measure sensitivity in the B/C ratio. The coefficient of determination ($R^2$) is calculated for each variable used in the B/C calculation. The $R^2$ measures the proportion of the variance associated with each independent variable. It is used to rank the effect of each independent variable on the B/C outcome.

$CO_2$ sequestration Cost

The quantity of atmospheric $CO_2$ captured is calculated using the methodology described by Brown et. al. (nd). The formula used to calculate the quantity sequestered is:

$$\frac{MT \text{ of biochar}}{MT \text{ of feedstock}} \times \frac{MT \text{ C}}{MT \text{ biochar}} \times \frac{44 \text{ MT CO2}}{12 \text{ MT C}} = MT \text{ CO2}$$

The biochar sequestration cost formula was developed by Timmons et.al. (2017) and is used to calculate the $CO_2$ sequestration cost.

$$Biochar\ sequestration\ cost = \frac{(K\alpha + C)}{\Delta CO_2} - B_a - B_c$$

Where:

K = capital cost of the biochar system
C = operating cost for the biochar system
$\Delta CO_2$ = change in atmospheric $CO_2$ which equals the amount of $CO_2$ sequestered.
Ba = the biochar benefit in agricultural use
Bc = the benefit of biochar coproducts

α = capital cost recovery factor:

$$\alpha = \frac{r(1+r)^T}{(1+r)^T - 1}$$

Where:

r = the discount rate
T = the useful life of the capital equipment

## Results and Discussion

As mentioned earlier, the model treats the wine industry value chain as a 'closed economy', that is, only wine industry pomace and grape prunings are used to produce the biochar and all the biochar production is used in the vineyards. This limits the amount of biochar that can be produced, and the vineyard area that can be treated. The mean values used in the model are: 3,500 tonnes of biochar is produced annually, 288 hectares of vineyard are treated with biochar, and 227,547 litres of additional wine is produced.

This section is divided into four parts: the biochar production sector, the vineyard sector, the winery sector, and carbon sequestration.

### The Benefits and Costs of Producing Biochar from Wine Industry Waste

The risk associated with biochar production is captured by the benefit cost analysis (B/C ratio). The project is considered viable if the B/C ratio > 1.0. The uncertainty ranges for each variable are provided in appendix B. The mean values from the Monte Carlo simulation are shown in table 1.

Table 1: Variables used in the Biochar Sector Benefit Cost Simulation

|  | Mean Value | |
| --- | --- | --- |
| Variable | Independent | Integrated |
| Production (tonnes) | 3,500 | 3,500 |
| Biochar price ($/tonne) | 1,077.67 | 1,077.67 |
| Variable cost ($/tonne) | 525.97 | 445.51 |
| Fixed cost | 894,120 | 355,760 |
| Capital cost | 493,990 | 383,982 |

The mean operating cost/tonne is $781.36 for the independent sector scenario and $455.56 for the integrated sector scenario. The lower operating costs for the integrated scenario reflect the synergies created through integration. The independent scenario cost is consistent with costs reported for sawmill waste by Clearly (nd). Clearly reported operating costs ranging from $515 to $824/tonne.

### Independent Biochar Scenario

The B/C ratio for the independent scenario ranged from 0.49 – 2.43, with a mean of 1.34. The probability that the B/C ratio will be greater than 1.0 is 79.9%. The benefit cost probability distribution is provided in figure 4.

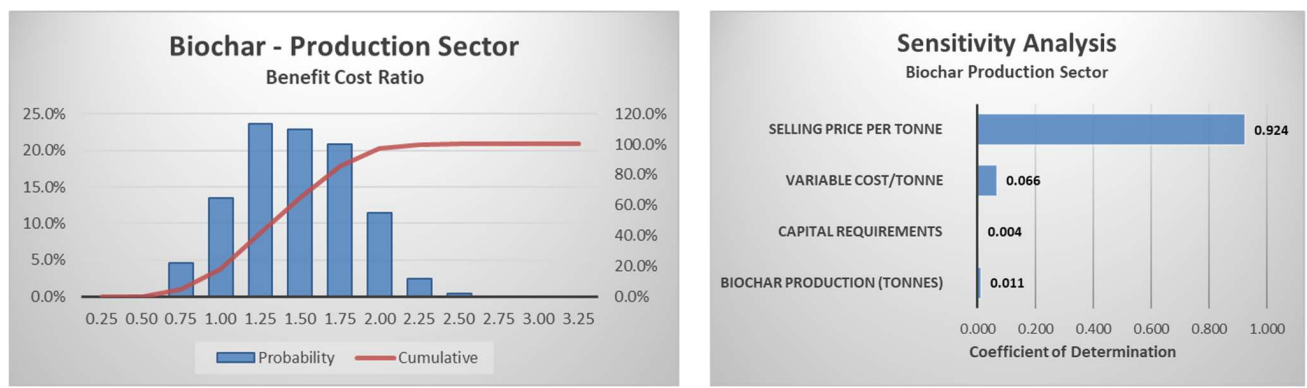

Figure 4: Independent Biochar Sector Benefit Cost Ration Probability Distribution and Sensitivity Analysis

A sensitivity analysis was completed for four variables: selling price per tonne, variable cost/tonne, capital requirements, and biochar production (tonnes). The results are shown in figure 4. The B/C ratio is most sensitive to the biochar price, which accounts for 92.4% of the variance in the B/C ratio. These findings are consistent with sensitivity analysis reported by Campbell et. al. (2018). All other variable showed limited impact on the B/C ratio.

The break-even price for biochar is $820.92. This is consistent with break-even values reported by Campbell et. al. (2018) that ranged from $838/tonne to $1,500/tonne.

The NPV calculations for the value chain are summarized in tables 9 and 10. The NPV range for the biochar sector is -$9,41 million to $22.73 million with a mean NPV of $5.87 million.

The range of NPV for the biochar sector is consistent with other results reported in the literature. Campbell et. al. (2018) reported a NPV range of -$US34 million to $US139 million with a mean of US$45 million. Their study was based on a much larger biochar production of 17,700 tonnes (five times larger). Haeldermans et. al. (2020) looked at six different feedstocks and reported that all NPVs, except for one were positive. The NPV ranged from a mean of €-1.77 million to €32.85 million

*Integrated Biochar Scenario*

The B/C ratio for the integrated scenario ranged from 0.89 – 4.40, with a mean of 2.34. The probability that the B/C ratio will be greater than 1.0 is 99.3%. The benefit cost probability distribution is provided in figure 5.

Consistent with the separate biochar scenario, the B/C ratio is most sensitive to the biochar price, which accounts for 92.4% of the variance in the B/C ratio.

The break-even price for the integrated scenario is $502.10. This is $318.82 lower than for the separate scenario. The NPV calculations for the value chain are summarized in tables 9 and 10. The NPV range for the biochar sector is -$1.26 million to $33.1 million with a mean NPV of $13.28 million.

The higher B/C ratio and NPVs for the integrated sector reflects the lower operating and capital requirements created through integration. The analysis indicates that these savings are significant.

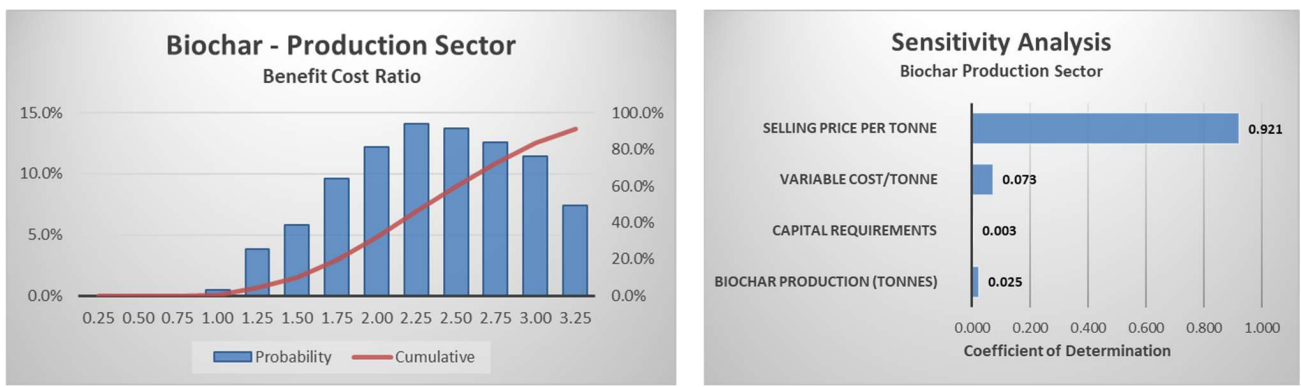

Figure 5: Integrated Biochar Sector Benefit Cost Ration Probability Distribution and Sensitivity Analysis

## The Benefits and costs of Using Biochar as a Soil Amendment in the Vineyard

The uncertainty ranges for each variable are provided in appendix B. The mean values from the Monte Carlo simulation are shown in table 2. The revenue and cost increases reflect the increase in grape yield/ha.

Table 2: Variables Used in the Vineyard Sector Benefit Cost Simulation

|  | Mean Value | |
| --- | --- | --- |
| Variable | Independent | Integrated |
| Area treated (hectares) | 288 | 288 |
| Grape yield increase | 15% | 15% |
| Grape revenue increase | 2,931.30 | 2,933.10 |
| Variable cost increase ($/ha) | 2,286.47 | 2,290.41 |
| Capital cost increase (planting) | 329,321 | 329,114 |

*Independent Biochar Scenario*

The B/C ratio for grape production ranged from 0.82 – 1.57, with a mean of 1.19. The probability that the B/C ratio will be greater than 1.0 is 91.2%. The benefit cost probability distribution is provided in figure 6.

A sensitivity analysis was completed for five variables: area treated (in Ha), grape price/tonne, revenue increase/Ha, variable cost increase/Ha, and capital requirements (planting costs). The results are shown in figure 6. The B/C ratio is most sensitive to the increased planting coats, which accounts for 27.4% of the variance in the B/C ratio, followed by variable cost increase of 18.9%, revenue increase at 18.6% and finally, area treated at 17.7%

The mean biochar application rate is 13.0 tonnes/ha. Using a biochar price of $1,077.67/tonne the application cost is $14,009.71/ha or $3,502.43/ha when amortized over four years, the useful life of the application.

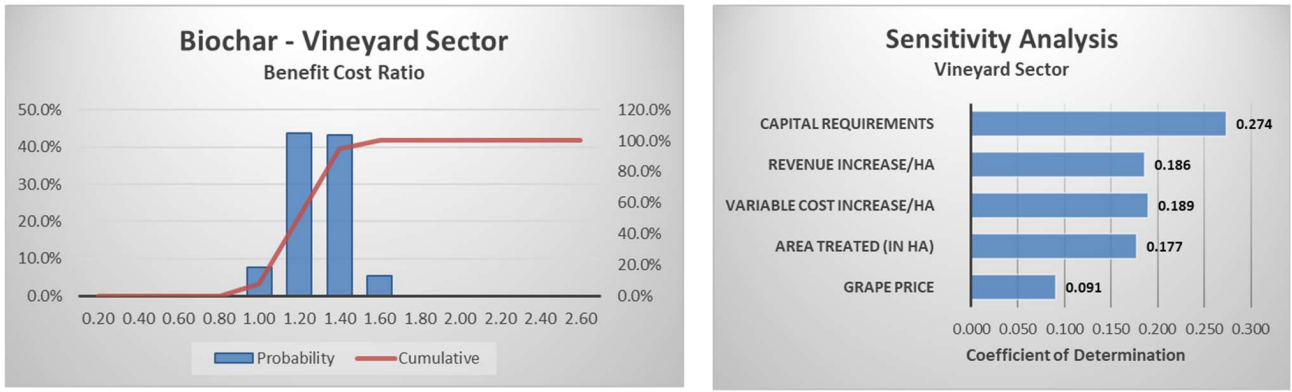

Figure 6: Vineyard Sector Benefit Cost Ratio Probability Distribution and Sensitivity Analysis Under the Independent Biochar Scenario

The increase in annual net income from the treated area range from -$90,780 to $899,865 with a mean of $195,812, or $679.90/hectare.

The NPV for the vineyard sector are shown in tables 9 and 10. The range is from -$825,405 to $4.63 million with a mean NPV of $ 873,859.

*Integrated Biochar Scenario*

The B/C ratio for grape production ranged from 0.81 – 1.62, with a mean of 1.19. The probability that the B/C ratio will be greater than 1.0 is 93.1%. The benefit cost probability distribution is provided in figure 7.

A sensitivity analysis was completed for five variables: area treated (in Ha), grape price/tonne, revenue increase/Ha, variable cost increase/Ha, and capital requirements (planting costs). The results are shown in figure 7. The B/C ratio is most sensitive to the increased planting coats, which accounts for 24.3% of the variance in the B/C ratio, followed by variable cost increase of 20.0%%, area treated at 15.8%. and finally, revenue increase at 14.3%

The mean biochar application rate is 13.0 tonnes/ha. Using a biochar price of $502.10/tonne the application cost is $6,527.30/ha or $1,631.83/ha when amortized over four years, the useful life of the application.

The increase in annual net income from the treated area range from -$87,667 to $741,717with a mean of $194,189, or $674.27/hectare.

The NPV for the vineyard sector are shown in tables 9 and 10. The range is from -$720,154 to $3.83 million with a mean NPV of $864,725.

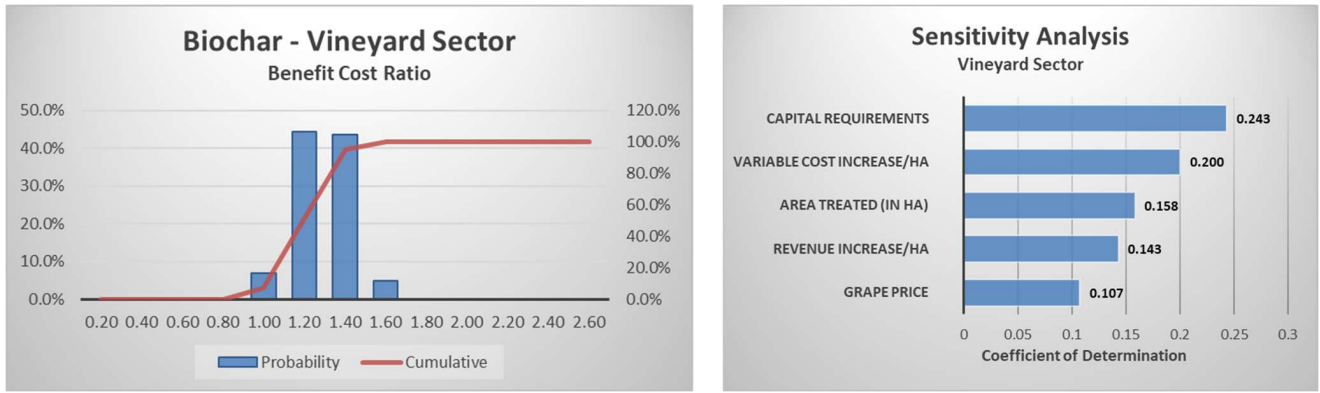

Figure 7: Vineyard Sector Benefit Cost Ratio Probability Distribution and Sensitivity Analysis Under the Integrated Biochar Scenario

## The Benefit to the Winery of Using Biochar in the Vineyard

The uncertainty ranges for each variable are provided in appendix B. The mean values from the Monte Carlo simulation are shown in table 3. The results were the same for both scenarios. The revenue and cost increases reflect the wine produced from the increase in grape yield/ha.

Table 3: Variables Used in the Winery Sector Benefit Cost Simulation

| Variable | Mean Value |
| --- | --- |
| Wine production increase (L) | 227,547 |
| Wine price to the winery ($/L) | 9.68 |
| Variable cost ($/L) | 5.97 |

The B/C ratio for wine production is greater than 1.0 and ranged from 1.45 to 1.81, with a mean of 1.62. The probability that the B/C ratio will be greater than 1.0 is 99.8%. The benefit cost probability distribution is provided in figure 8.

A sensitivity analysis was completed for three variables: wine price ($/L), wine production increase (L), and variable cost ($/L). The results are shown in figure 8. The B/C ratio is most sensitive to the price of wine, which accounts for 99.99% of the variance in the B/C ratio. All other variable showed limited impact on the B/C ratio.

The increase in annual net income from the addition wine production ranged from $290,587 to $2.16 million with a mean of $843,317.

The NPV for the winery sector is shown in tables 9 and 10. The range is from $1.75 million to $12.29 million with a mean of $5.18 million.

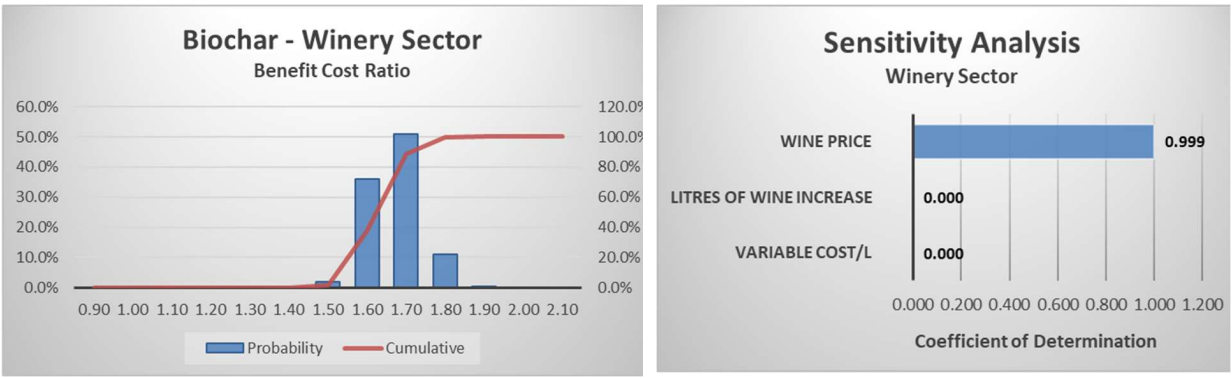

Figure 8: Winery Sector Benefit Cost Ratio Probability Distribution and Sensitivity Analysis

## Atmospheric $CO_2$ Sequestered by Producing Biochar and Using it as a Soil Amendment in the Vineyard

The quantity of atmospheric $CO_2$ sequestered by the wine industry and the sequestration cost/tonne are calculated for each scenario. The mean values for the variables used in the Monte Carlo simulation and used for these calculations are shown in table 4.

Table 4: Variables Used in the Carbon Sequestration Calculations

|  | Mean Value | |
| --- | --- | --- |
| Variable | Independent | Integrated |
| Grape prunnings used (tonnes) | 7,107 | 7,107 |
| Pomace used (tonnes) | 3,556 | 3556 |
| Biochar conversion rate | 32% | 31% |
| Biochar carbon content | 70% | 70% |
| Carbon conversion to $CO_2$ | 3.67 | 3.67 |
| Capital recovery factor | 0.117 | 0.117 |

The quantity of $CO_2$ sequestered is the same under each scenario and ranged from 0 to 10,800 tonnes with a mean of 8,990 tonnes, or 31.2 tonnes per treated hectare.

### Carbon Offset Program

A carbon offset program is available in British Columbia. $CO_2$ emitters such as airlines or oil and gas producers can purchase carbon offsets to reduce their carbon footprint. Carbon offset programs provide a cash payment to parties that can verify that they have sequestered $CO_2$, and provide governments with a tool to meet their carbon reduction targets. In BC, Independent validators and verifiers provide third-party reviews to ensure that the offsets are verifiable and incremental. The Ministry of Environment provides regulatory oversight.

Carbon offsets can provide an additional source of revenue for grape growers. There is no set price for the offsets, and buyers and sellers negotiate a price. Carbon offset sellers must therefore know their carbon sequestration cost. The carbon sequestration costs for each scenario are provided in table 5.

Table 5: Carbon Sequestration Costs Under Each Scenario

|  | Independent | Integrated |
|---|---|---|
| *Carbon Sequestration Cost* | | |
| Maximum cost ($/t) | 78.95 | 61.55 |
| Mean cost ($/t) | 62.37 | 47.64 |
| Maximum cost ($/t) | 0.00 | 0.00 |

In this study, carbon sequestration costs under the independent scenario range from zero to $78.95/tonne with a mean of $62.37/tonne. Under the integrated scenario they range from zero to $61.55/tonne with a mean of $47.64/tonne. The lower sequestration cost for the integrated scenario is due to the lower biochar production cost. In comparison, Timmons et.al (2017) reported carbon sequestration costs in the range of range $82 to $114/tonne. Gillingham and Stock (2018) reported estimate $CO_2$ abatement costs ranging from $32 - $95 using different carbon capture and storage technologies. The additional revenue associated with carbon offsets by grape growers is provided further below in table 8.

## Wine and Biochar Value Chain Summary

The contribution of each sector to the wine and biochar value chain is summarized in table 6 and table 7.

Table 6: Wine and Biochar Value Chain Summary Under the Independent Scenario

|  | Biochar | Vineyard | Winery | Value Chain Total |
|---|---|---|---|---|
| Annual Net Income | 1,035,934 | 195,812 | 844,981 | 3,294,919 |
| NPV | 5,871,360 | 873,859 | 5,192,042 | 19,051,842 |
| Annual Net Income/Ha | 3,595 | 679 | 2,932 | 11,433 |
| NPV per Ha | 20,374 | 3,032 | 25,381 | 57,791 |
| Benefit Cost Ratio | 1.34 | 1.19 | 1.62 |  |
| Probability of B/C > 1.0 | 79.9% | 91.2% | 99.8% |  |

Table 7: Wine and Biochar Value Chain Summary Under the Integrated Scenario

|  | Biochar | Vineyard | Winery | Value Chain Total |
|---|---|---|---|---|
| Annual Net Income | 2,217,005 | 194,189 | 840,586 | 3,251,780 |
| NPV | 13,238,563 | 864,725 | 5,165,034 | 19,268,323 |
| Annual Net Income/Ha | 7,699 | 674 | 2,919 | 11,292 |
| NPV per Ha | 45,972 | 3,003 | 27,364 | 76,338 |
| Benefit Cost Ratio | 2.34 | 1.19 | 1.62 |  |
| Probability of B/C > 1.0 | 99.3% | 93.1% | 99.8% |  |

The carbon sequestration results are summarized in table 6.

Table 8: Carbon Sequestration and Carbon Offset Benefit

|  | Independent | Integrated |
|---|---|---|
| *Carbon Sequestration* | | |
| Mean carbon sequestered annually (t) | 8,976 | 9,001 |
| | | |
| *Potential Carbon Offset Benefit* | | |
| Mean carbon offset benefit ($/t) | 62.37 | 47.64 |
| Mean carbon offset benefit ($/Ha) | 807.04 | 618.53 |
| Total carbon offset benefit ($) | 232,579 | 178,121 |

The range NPV values for the wine and biochar value chain under scenario are summarized in table 9 and table 10.

Table 9: NPV Calculations for each Value Chain Sector Under the Independent Scenario

|  | Minimum | Mean | Maximum |
|---|---|---|---|
| Biochar Production Sector | -9,416,314 | 5,871,360 | 22,730,947 |
| Vineyard Sector | -825,405 | 873,859 | 4,630,695 |
| Winery Sector | 2,271,514 | 5,192,042 | 15,967,066 |
|  | -7,970,204 | 11,937,262 | 43,328,709 |

Table 10: NPV Calculations for each Value Chain Sector Under the Integrated Scenario

|  | Minimum | Mean | Maximum |
|---|---|---|---|
| Biochar Production Sector | -1,265,853 | 13,238,563 | 33,092,783 |
| Vineyard Sector | -720,154 | 864,725 | 3,827,397 |
| Winery Sector | 2,044,217 | 5,165,034 | 13,270,016 |
|  | 58,210 | 19,268,323 | 50,190,196 |

## Conclusions and Recommendations

Under both scenarios, the addition of a biochar sector to the BC wine industry can increase the profitability of the industry while helping the industry to achieve its sustainability goals. An additional annual net income of $3.2 million is possible, a return of $2.81 for each dollar invested (1.19 for vineyards and 1.62 for wineries); however, this is not without risk. Under the independent biochar scenario, the probability of success ranges from 80% for the biochar sector to 99% for the winery sector. The high risk for the biochar and vineyard sectors is due to level of uncertainty associated with biochar production, and the increased yield in the vineyard. The level of risk in the vineyard sector is reduced under the integrated biochar scenario due to a lower biochar production cost. The probability of success ranges from 91% for the vineyard sector to 99.8% for the biochar and winery sectors.  Further research is required to remove some of uncertainty associated with the assumptions used in the models.

Biochar is a soil amendment, not a fertilizer, and it does not replace the use of nitrogen fertilizer. The high application rates of biochar and its high cost are a real barrier to its adoption, especially by smaller vineyards. Moreover, it is unknown how frequently biochar must be applied. The authors of the research cited in this study agree that biochar need not be applied each year, however there is no agreement regarding the frequency. To reduce the high application cost/ha some sources suggest that rather than a single application, biochar might be applied at a lower rate over several years, thus improving cash flows. Further research is needed to confirm this.

The risk associated with the uncertainty can be dealt with by conducting further research, but the high cost of biochar to the grape grower needs to be addressed. For the biochar production segment to be economically viable for an independent producer, the break-even price for biochar is $820/tonne, resulting in an application cost $10,600/ha, although this cost could be amortized over several years. However, when biochar production is integrated with the vineyard, the biochar cost/ha is reduced to $517/tonne resulting in an application cost of $ 6,700/ha or $1,675/ha when amortized over four years, and $1,340 over five years. As a comparison, the cost of ammonium nitrate fertilizer is $300/ha[4], or $1,200/ha when applied annually for four years.

Biochar production from grape prunings and pomace is an effective way to sequester atmospheric $CO_2$. Results show that even with biochar production of only 3,500 tonnes, 9,000 tonnes of $CO_2$ can be sequestered annually, the equivalent of removing 2,100 cars from the road (United States Environmental Protection Agency, nd). Also, additional revenue from carbon offsets may be available to the grape grower.

In this study, the value chain is considered to be a closed economy, with the quantity of biochar production limited by the quantity of wine industry waste. However, the technology can be readily scaled up to produce much greater quantities of biochar. This can be accomplished using other sources of biomass for production. For example, in the central Okanagan, the Kelowna landfill receives 56,000 tonnes of residential yard waste, enough to produce 20,000 tonnes of biochar, enough to treat up to 2,000 hectares of vineyard and sequester 50,000 tonnes of $CO_2$.

*Recommendations*

Of the two scenarios investigated in this study, the integration option provides the most attractive way to commercialize biochar production. The risk is lower than for an independent producer and there is a significant cost advantage available to the integrated vineyard, $517/tonne compared to $1,077/tonne when purchased from a reseller. Also, due to the high biochar retail price, it is unlikely that a new independent biochar producer would be successful.

The maximum benefit of using biochar will likely occur with new grape plantings. When biochar is added to the soil prior to planning the grapes, prior research indicates that the young vines may develop better root systems, be more tolerant to drought and have higher yields than new plantings that are not treated with biochar (Amendola, et.al.; Baronti et.al).

A majority of Canadians believe that climate change is a problem. Recent research by Abacus Data (2019) found that 82% say climate change is a serious problem, with 47% describing it as an extremely serious problem. British Columbians are among the most concerned Canadians, with 86% believing that it is a serious problem compared to 69% of Albertans. Moreover, younger Canadians are more concerned about climate change than older age groups. Of those aged 18 – 29 years, 65% believe that there is a climate emergency compared to 59% for those over 45 years. This presents an opportunity for the BC wine industry to differentiate itself from foreign wines by branding BC wines as 'climate friendly'. BC wine consumers can enjoy a glass of premium BC wine while feeling good about climate change mitigation. Beverage manufacturers are already moving in this direction. Anheuser-Busch InBev plans to brew Michelob Ultra Pure Gold beer using solar energy; the goal is to appeal to younger, more environmentally conscious consumers (Hirtzer, M. 2012).

---

[4] Adapted from the Alberta Fertilizer Guide. Alberta Agriculture, Food and Rural Development, https://www1.agric.gov.ab.ca/$department/deptdocs.nsf/all/agdex3894/$file/541-1.pdf?OpenElement

Further research is necessary to deal with the uncertainty inherent in this study. The next stage for research should include the following:

1. Complete primary research to understand what wine industry stakeholders know about biochar and its use. This can be done using focus groups and surveys.
2. Construct a prototype biochar production facility to verify the biomass conversion rates and the production costs.
3. Establish vineyard test plots to verify: the application rates, application frequency, and potential increase in grape yields. This needs to be a longitudinal study conducted over five years.

Ellen MacArthur Foundation (nd). What is a circular economy? A framework for an economy that is restorative and regenerative by design. Retrieved from https://www.ellenmacarthurfoundation.org/circular-economy/concept

Farm Folk City Folk (nd). Biochar Manual for Small Farms in BC. Retrieved from https://sites.google.com/site/fcfcbiocharmanual

Gasser, T., Guivarch, C., Tachiiri, K., Jones, C.C. & Ciais, P. (2015). Negative emissions physically needed to keep global warming below 2 °C. Nature Communications 6, 7958 (2015). https://doi.org/10.1038/ncomms8958

Genesio, L., Miglietta, F., Baronti, S. and Vaccari, F. P. (2015). Biochar increases vineyard productivity without affecting grape quality: Results from a four years field experiment in Tuscany. Agriculture, Ecosystems & Environment 201:20-25.

Giagnoni, L., Maienza, A., Baronti, S., Vaccari, F. P., Genesio, L., Taiti, C., Martellini, T., Scodellini, R., Cincinelli, A., Costa, C. and others. (2019). Long-term soil biological fertility, volatile organic compounds and chemical properties in a vineyard soil after biochar amendment. Geoderma 344:127-136.

Gillingham, K. & Stock, J. (2018). The Cost of Reducing Greenhouse Gas Emissions. Forthcoming, Journal of Economic Perspectives. Retrieved from https://scholar.harvard.edu/files/stock/files/gillingham_stock_cost_080218_posted.pdf

Government of Canada a (nda) Net-Zero Emissions by 2050 https://www.canada.ca/en/services/environment/weather/climatechange/climate-plan/net-zero-emissions-2050.html

Government of Canada (ndb) Crude Oil Facts. Retrieved from https://www.nrcan.gc.ca/science-data/data-analysis/energy-data-analysis/energy-facts/crude-oil-facts/20064

Government of Canada (ndc) New regulations: Guide to submitting applications for registration under the Fertilizers Act. Retrieved from https://www.inspection.gc.ca/plant-health/fertilizers/registering-fertilizers-and-supplements/new-regulations-guide/eng/1601391948941/1601392244498?chap=0

Haeldermansa, T., Campionc, L., Kuppensc, T., Vanreppelena, K,. Cuypersd, A,. & Schreurs, S. (2020). A comparative techno-economic assessment of biochar production from different residue streams using conventional and microwave pyrolysis. Bioresource Technology 318 (2020) 124083. doi.org/10.1016/j.biortech.2020.124083

Hepburn, C., Adlen, E., Beddington, J., Carter, E., Fuss, S., Mac Dowell, N., Minx, J., Smith, P., Williams, C. (2019). The technological and economic prospects for $CO_2$ utilization and removal. Nature, 575: 87-95. doi: 10.1038/s41586-019-1681-6

Hirtzer, M. AB InBev Tries to Sell a Better Climate with Solar-Powered Beer. *Bloomberg Green*. Retrieved from https://www.bloomberg.com/news/articles/2021-04-19/ab-inbev-tries-to-sell-a-better-climate-with-solar-powered-beer
23

Hogervorst, J. C., Miljić U, & Puškaš, V. (2017). 5 - Extraction of Bioactive Compounds from Grape Processing By-Products. Handbook of Grape Processing By-Products. Academic Press, Pages 105-135. ISBN 9780128098707. https://doi.org/10.1016/B978-0-12-809870-7.00005-3.

Keske, C. (2020). Biochar: an emerging market solution for legacy mine reclamation and the environment. Appalachian Natural Resources Law Journal, 6, 1-14

Khorram, M. S., Zhang, G., Fatemi, A., Kiefer, R., Maddah, K., Baqar, M., Zakaria, M. P. and Li, G. (2019). Impact of biochar and compost amendment on soil quality, growth and yield of a replanted apple orchard in a 4-year field study. Journal of the Science of Food and Agriculture 99(4):1862-1869.

Major, J. (2010). Guidelines on Practical Aspects of Biochar Application to Field Soils in Various Soil Management Systems. International Biochar Initiative.

McKinsey Centre for Business and Environment (nd). Growth Within: A Circular Economy Vision for A Competitive Europe. Retrieved from https://www.ellenmacarthurfoundation.org/assets/downloads/publications/EllenMacArthurFoundation_Growth-Within_July15.pdf

Pembina Institute (2010). Greenhouse Gas Sources and Sinks in Canada 1990–2008. Adapted by the Pembina Institute from Environment Canada, National Inventory Report, Ottawa, ON: Environment Canada, 2010), Part 1: 86, 89–90; Part 3: 4. 2

Rawat, J., Saxena , J., & Sanwal, P. (2019). Biochar: a sustainable approach for improving plant growth and soil properties. Biochar: A Sustainable Approach for Improving Plant Growth and Soil Properties. Retreived from http://dx.doi.org/10.5772/intechopen.82151

Sahoo, K., Bilek, E., Bergman, R., & Mani, S. (2019). Techno-economic analysis of producing solid biofuels and biochar from T forest residues using portable systems. Applied Energy, 235 578 – 590. Retrieved from https://doi.org/10.1016/j.apenergy.2018.10.076

Skinkis, P. (2013). How to Measure Dormant Pruning Weight of Grapevines. Oregon State University. Retrieved from https://ir.library.oregonstate.edu/downloads/3x816m914.

Timmons, D,. Lema-Driscoll, A., & Gazi Uddin, G- (2017). The Economics of Biochar Carbon Sequestration in Massachusetts. University of Massachusetts Boston.

United Nations (nd). Retrieved from https://unfccc.int/climate-action/momentum-for-change/climate-neutral-now/carbon-neutral-government-program-canada

United States Environmental Protection Agency. (nd). Greenhouse Gas Emissions from a Typical Passenger Vehicle. Retrieved from https://www.epa.gov/greenvehicles/greenhouse-gas-emissions-typical-passenger-vehicle#:~:text=A%20typical%20passenger%20vehicle%20emits%20about%204.6%20metric,questions%20about%20greenhouse%20gas%20emissions%20from%20passenger%20vehicles
24

## Appendix A: $CO_2$ Utilization Pathways

Ten $CO_2$ utilization and removal pathways

| Pathway[a] | Removal and/or capture[b] | Utilization product | Storage[c,d] and likelihood of release (high/low) | Emission on use[f] or release during storage[g] | Example cycles[h] |
|---|---|---|---|---|---|
| (1) Chemicals from $CO_2$ | Catalytic chemical conversion of $CO_2$ from flue gas or other sources into chemical products | $CO_2$-derived platform chemicals such as methanol, urea and plastics | Various chemicals (days/decades) – high | Hydrolysis or decomposition | KCLG; KCLF; ALFJ; ALG |
| (2) Fuels from $CO_2$ | Catalytic hydrogenation processes to convert $CO_2$ from flue gas or other sources into fuels | $CO_2$-derived fuels such as methanol, methane and Fischer–Tropsch-derived fuels | Various fuels (weeks/months) – high | Combustion | KCLG; ALG |
| (3) Products from microalgae | Uptake of $CO_2$ from the atmosphere or other sources by microalgae biomass | Biofuels, biomass, or bioproducts such as aquaculture feed | Various products (weeks/months) – high | Combustion (fuel) or consumption (bioproduct) | KCLG; BG |
| (4) Concrete building materials | Mineralization of $CO_2$ from flue gas or other sources into industrial waste materials, and $CO_2$ curing of concrete | Carbonated aggregates or concrete products | Carbonates (centuries) – low | Extreme acid conditions | KCLF; ALF |
| (5) $CO_2$-EOR | Injection of $CO_2$ from flue gas or other sources into oil reservoirs | Oil | Geological sequestration (millennia) – low[e] | n.a. | KCD |
| (6) Bioenergy with carbon capture and storage (BECCS) | Growth of plant biomass | Bioenergy crop biomass | Geological sequestration (millennia) – low[e] | n.a. | BCD |
| (7) Enhanced weathering | Mineralization of atmospheric $CO_2$ via the application of pulverized silicate rock to cropland, grassland and forests | Agricultural crop biomass | Aqueous carbonate (centuries) – low | Extreme acidic conditions | BE |
| (8) Forestry techniques | Growth of woody biomass via afforestation, reforestation or sustainable forest management | Standing biomass, wood products | Standing forests and long-lived wood products (decades to centuries) – high | Disturbance, combustion or decomposition | BFJ |
| (9) Soil carbon sequestration techniques | Increase in soil organic carbon content via various land management practices | Agricultural crop biomass | Soil organic carbon (years to decades) – high | Disturbance or decomposition | BFJ |
| (10) Biochar | Growth of plant biomass for pyrolysis and application of char to soils | Agricultural or bioenergy crop biomass | Black carbon (years to decades) – high | Decomposition | BFJ |

n.a., not applicable.
[a]The ten pathways are depicted in Fig. 1 and are represented as a combination of steps in Fig. 2.
[b]Removal and/or capture corresponds to steps A, B and/or C in Fig. 2.
[c]Storage corresponds to steps D, E or F in Fig. 2.
[d]Storage durations represent best-case scenarios. For instance, in $CO_2$-EOR, if the well is operated with complete recycle, the $CO_2$ is trapped and can be stored on a timescale of centuries or more[22]. This is also relevant only for conventional operations.
[e]Release during geological storage is usually a consequence of engineering implementation error.
[f]Emission on use corresponds to step G in Fig. 2.
[g]Release during storage corresponds to steps H, I or J in Fig. 2.
[h]The letters stated are the steps from Fig. 2 that comprise the example cycle.

Source: Hepburn, C., Adlen, E., Beddington, J., Carter, E., Fuss, S., Mac Dowell, N., Minx, J., Smith, P., Williams, C. (2019)

## Appendix B: Monte Carlo Uncertainty Distributions

### Independent Biochar Sector

| Variable | Lowest | Base | Highest |
|---|---|---|---|
| Pomace supply (tonnes) | 3,556 | 3,556 | 3,556 |
| Prunings supply (tonnes) | 5,991 | 7,107 | 8,222 |
| Pomace Cost ($/t) | 0.00 | 5.00 | 10.00 |
| Prunings Cost ($/t) | 0.00 | 10.00 | 40.00 |
| Biochar conversion Rate (%) | 0.25 | 0.33 | 0.40 |
| Biochar price ($/t) | 334 | 1,078 | 1,822 |
| *Biochar Cost* | | | |
| Fixed cost ($/t) | 267.73 | 267.73 | 267.73 |
| Variable cost | 405.95 | 487.14 | 649.51 |
| Capital equipment | 320,600 | 475,200 | 742,500 |

### Integrated Biochar Sector

| Variable | Lowest | Base | Highest |
|---|---|---|---|
| Pomace supply (tonnes) | 3,556 | 3,556 | 3,556 |
| Prunings supply (tonnes) | 5,991 | 7,107 | 8,222 |
| Pomace Cost ($/t) | 0.00 | 5.00 | 10.00 |
| Prunings Cost ($/t) | 0.00 | 10.00 | 40.00 |
| Biochar conversion Rate (%) | 0.25 | 0.33 | 0.40 |
| Biochar price ($/t) | 334 | 1,078 | 1,822 |
| *Biochar Cost* | | | |
| Fixed cost ($/t) | 99.35 | 99.35 | 99.35 |
| Variable cost | 278.04 | 333.65 | 444.87 |
| Capital equipment | 230,600 | 365,200 | 612,500 |

*Vineyard Sector*

| Variable | Lowest | Base | Highest |
|---|---|---|---|
| Total grape production (Tonnes) | 70,874 | 80,292 | 95,720 |
| Total grapes (Tonnes/Ha)) | 6.91 | 7.83 | 9.33 |
| Yield increase (%) | 0.10 | 0.15 | 0.20 |
| Average grape price ($/t) | 2,227 | 2,451 | 2,675 |
| Average grape direct cost ($Ha) | 13,642 | 13,642 | 13,642 |
| Average grape capital cost ($/t) | 955 | 955 | 955 |
| Biochar application rate (t/Ha) | 5.00 | 12.75 | 22.00 |
| Biochar price ($/t) | 333.94 | 1,077.74 | 1,821.53 |
| Biochar cost per Ha | 1,669.70 | 13,741.15 | 40,073.77 |
| Application frequency (Per Year) | 0.14 | 0.25 | 0.50 |
| Application frequency (Years) | 7 | 4 | 2 |
| % of total hectares treated | 0.05 | 0.10 | 0.15 |

*Winery Sector*

| Variable | Lowest | Base | Highest |
|---|---|---|---|
| White wine price to winery ($/L) | 7.35 | 8.60 | 9.80 |
| Red wine price to winery ($/L) | 9.50 | 10.74 | 12.49 |
| White wine cost ($/L) | 5.61 | 5.61 | 5.61 |
| Red wine cost ($/L) | 6.35 | 6.35 | 6.35 |
| White wine percent of crop | 0.49 | 0.51 | 0.54 |
| Red wine percent of crop | 0.51 | 0.49 | 0.46 |
| Grape yield (Tonnes/Ha) | 1.38 | 1.38 | 1.38 |
| White wine (Litres/Ha) | 463.38 | 463.38 | 463.38 |
| Red wine (Litres/Ha) | 445.65 | 445.65 | 445.65 |
| White wine revenue ($/Ha) | 3,405.87 | 3,983.20 | 4,541.16 |
| Red wine revenue ($/Ha) | 4,231.66 | 4,788.45 | 5,567.97 |
| Total wine revenue ($/Ha) | 7,637.52 | 8,771.65 | 10,109.13 |
| Total wine revenue ($/L) | 8.40 | 9.65 | 11.12 |
| Total wine cost ($/Ha) | 5,429.95 | 5,429.95 | 5,429.95 |
| Total wine cost($/L) | 5.97 | 5.97 | 5.97 |
| Hectares treated with biochar | 288 | 288 | 288 |